\documentstyle[12pt,graphicx]{article}

\begin{document}

\title{SOME PROPERTIES OF A NEW SOLUTION OF THE ERNST EQUATION}
\author{J. Gariel$^{1}$\thanks{%
e-mail: gariel@ccr.jussieu.fr}, G. Marcilhacy$^{1}$ and N. O. Santos$%
^{1,2,3} $\thanks{%
e-mail: nos@cbpf.br} \\
\\
{\small $^{1}$Universit\'{e} Pierre et Marie Curie - CNRS/UMR 8540,}\\
{\small LRM/ERGA, Tour 22-12, 4\'eme \'etage, Bo\^{\i}te 142, 4 place
Jussieu,}\\
{\small 75252 Paris Cedex 05, France.}\\
{\small $^{2}$Laborat\'{o}rio Nacional de Computa\c{c}\~{a}o Cient\'{i}fica,}%
\\
{\small 25651-070 Petr\'opolis RJ, Brazil.}\\
{\small $^{3}$Centro Brasileiro de Pesquisas F\'{i}sicas}\\
{\small 22290-180 Rio de Janeiro RJ, Brazil.}}
\maketitle

\begin{abstract}
From a particularly simple solution of the Ernst equation, we build a
solution of the vacuum stationary axisymmetric Einstein equations depending
on three parameters. The parameters are associated to the total mass of the
source and its angular momentum. The third parameter produces a topological
deformation of the ergosphere making it a two-sheet surface, and for some of
its values forbids the Penrose process.
\end{abstract}

\section{Introduction}

We propose a new solution of the Ernst equation. This solution is symmetric
in the prolate spheroidal coordinates and depends on one parameter, called $%
q_{1}$. The construction of the corresponding gravitational potentials, with
the help of the Boyer-Linquist transformation, shows that this solution has
not a good asymptotical behaviour for its dragging. But, by an Ehlers
transformation followed by an unitary transformation, which introduces two
more parameters, an asymptotically flat solution can be easily built. So,
the obtained solution, in Boyer-Linquist coordinates, depends on three
parameters, which are connected, as in the Kerr case, to the total mass of
the source and its angular momentum. However, we did not succeed to relate
directly this solution to the Kerr solution, but we know that such a linck
does exist by reason of the uniqueness theorem for the solution with a good
asymptotical behaviour and without naked singularity \cite{Carter}. Only the
extreme black hole of Kerr appears as a limit of the proposed solution, and
this latter does not present a naked singularity. Varying the $q_{1}$
parameter allows to show its role in the ergosphere shape. The ergosphere,
which initially has a torus shape, continuously looses its form and finally
separates into a two-sheet toroidal surface, progressively exposing more and
more the event horizon. Then the Penrose process \cite{Penrose} is no longer
able to take place in a domain of the azimuthal angle, for some range of the 
$q_{1}$ parameter values.

\section{Brief recall on the resolution of the Ernst equation}

The line element of a general axisymmetric stationary spacetime is the so
called Papapetrou metric, which in the cylindrical coordinates, $\rho $, $z$
and $\phi $, reads 
\begin{equation}
ds^{2}=f(dt-\omega d\phi )^{2}-f^{-1}[e^{2\gamma }(d\rho ^{2}+dz^{2})+\rho
^{2}d\phi ^{2}],
\end{equation}
where the gravitational potentials, $f$, $\omega $ and $\gamma $ are
functions of $\rho $ and $z$ only. The canonical coordinates of Weyl, $\rho $
and $z$, can be given in terms of prolate spheroidal coordinates, $\lambda $
and $\mu $, by the relations 
\begin{equation}
\rho =k(\lambda ^{2}-1)^{1/2}(1-\mu ^{2})^{1/2},\;\;z=k\lambda \mu ,
\end{equation}
with 
\begin{equation}
\lambda \geq 1,\;\;-1\leq \mu \leq 1,\;\;k=\mbox{constant}.
\end{equation}
The metric (1) with relations (2) can be rewritten like 
\begin{eqnarray}
ds^{2} &=&f(dt-\omega d\phi )^{2}  \nonumber \\
&&-\frac{k^{2}}{f}\left[ e^{2\gamma }(\lambda ^{2}-\mu ^{2})\left( \frac{%
d\lambda ^{2}}{\lambda ^{2}-1}+\frac{d\mu ^{2}}{1-\mu ^{2}}\right) +(\lambda
^{2}-1)(1-\mu ^{2})d\phi ^{2}\right] ,
\end{eqnarray}
where the potentials are now functions of $\lambda $ and $\mu $. The Ernst
equation is \cite{Ernst} 
\begin{equation}
(\xi \bar{\xi}-1)\nabla ^{2}\xi =2\bar{\xi}\nabla \xi \cdot \nabla \xi ,
\end{equation}
where $\nabla $ and $\nabla ^{2}$ are the gradient and the three-dimensional
Laplacian operators respectively, $\bar{\xi}$ is the conjugated complex
potential of $\xi $, and in general its solution can be expressed as 
\begin{equation}
\xi (\lambda ,\mu )=P(\lambda ,\mu )+iQ(\lambda ,\mu ),
\end{equation}
where $P$ and $Q$ are real functions of $\lambda $ and $\mu $. Among the
classical solutions of the Ernst equation, we can cite the well known Kerr
solution \cite{Ernst}, 
\begin{equation}
\xi _{K}=p\lambda +iq\mu ,
\end{equation}
where $p$ nd $q$ are real constants satisfying 
\begin{equation}
p^{2}+q^{2}=1;
\end{equation}
and the Tomimatsu-Sato solution, 
\begin{equation}
\xi _{TS}=\frac{\alpha (\lambda ,\mu ;p,q,\delta )}{\beta (\lambda ,\mu
;p,q,\delta )},
\end{equation}
where $\alpha $ and $\beta $ are two complex polynomials depending on the
Kerr parameters $p$ and $q$ and a parameter $\delta $ assuming integer
values describing the deformation of the source \cite{Tomimatsu}. To
determine the potentials $f$, $\omega $ and $\gamma $ of the metric (4), the
method consists to use the following relation between $f$, the twist
potential $\Phi $ and $\xi $, 
\begin{equation}
f+i\Phi =\frac{\xi -1}{\xi +1},
\end{equation}
which implies with (6), 
\begin{equation}
f=\frac{P^{2}+Q^{2}-1}{R^{2}},\;\;\Phi =\frac{2Q}{R^{2}},
\end{equation}
or, equivalently, 
\begin{equation}
f=1-\frac{\partial }{\partial P}[\ln (R^{2}+Q^{2})],\;\;\Phi =\frac{\partial 
}{\partial Q}[\ln (R^{2}+Q^{2})],
\end{equation}
where 
\begin{equation}
R^{2}=(P+1)^{2}+Q^{2}.
\end{equation}
In (10), $\Phi $ is the twist potential defined up to a constant and related
to the dragging $\omega $ by the following differential equations, 
\begin{equation}
\frac{\partial \omega }{\partial \lambda }=\frac{k(1-\mu ^{2})}{f^{2}}\frac{%
\partial \Phi }{\partial \mu },\;\;\frac{\partial \omega }{\partial \mu }=-%
\frac{k(\lambda ^{2}-1)}{f^{2}}\frac{\partial \Phi }{\partial \lambda }.
\end{equation}
The potential $\omega $ is obtained by integration of (14), and $\gamma $ is
determined by quadratures. Any solution of the Ernst equation is a solution
of the Einstein equations.

\section{A particular new solution}

It is easy to verify that a particular rational solution of the Ernst
equation (5) is obtained when (6) has the following expressions for $%
P(\lambda ,\mu )$ and $Q(\lambda ,\mu )$, 
\begin{equation}
P=-q_{1}(\lambda +\mu ),\;\;Q=\frac{1+\lambda \mu }{\lambda +\mu },
\end{equation}
where $q_{1}$ is an arbitrary real parameter. In (15) we note a symmetry
between $\lambda $ and $\mu $. However, it can be proved, it has not an
axymptotically flat behaviour for the potentials $f$ and $\omega $. To
obtain a flat asymptotical behaviour, a first step is to introduce a second
real parameter, $\alpha _{1}$, by means of the following particular Ehlers
transformation \cite{Chandrasekhar} on (15), 
\begin{equation}
\xi _{1}=\frac{c_{1}\xi +d_{1}}{\bar{d}_{1}\xi +\bar{c}_{1}},
\end{equation}
where 
\begin{equation}
c_{1}=1+i\alpha _{1},\;\;d_{1}=i\alpha _{1},
\end{equation}
satisfying 
\begin{equation}
\left( 
\begin{array}{cc}
c_{1} & d_{1} \\ 
\bar{d}_{1} & \bar{c}_{1}
\end{array}
\right) \in SU(1,1),\;\;|c_{1}|^{2}-|d_{1}|^{2}=1.
\end{equation}
But, again, it can be proved, the solution (16) has still not the suitable
asymptotical flatness. Then, a second step consists to perform an unitary
transformation on $\xi _{1}$, 
\begin{equation}
\xi _{2}=e^{i\theta _{0}}\xi _{1}=(m+in)\xi _{1},\;\;m^{2}+n^{2}=1,
\end{equation}
with $\theta _{0}$ an arbitrary complex constant, and $m$ and $n$ real
constants. Then (19) with (16) and (17) becomes 
\begin{equation}
\xi _{2}=\frac{P(m-\alpha _{1}n)-Q(\alpha _{1}m+n)-\alpha _{1}n+i[P(\alpha
_{1}m+n)+Q(m-\alpha _{1}n)+\alpha _{1}m]}{-(\alpha _{1}Q+1)+i\alpha _{1}(P+1)%
}.
\end{equation}
Considering 
\begin{equation}
\alpha _{1}=-\frac{n}{2(1+m)}
\end{equation}
and applying the method recalled in section 2, we find the potentials
corresponding to the solution (20) of the Ernst equation 
\begin{eqnarray}
f &=&\left\{ \frac{(1+\lambda \mu )^{2}+(\lambda +\mu
)^{2}[q_{1}^{2}(\lambda +\mu )^{2}-1]}{(1+\lambda \mu )^{2}+(\lambda +\mu
)^{2}[q_{1}(\lambda +\mu )+1]^{2}]}\right\} \cos ^{-2}\frac{\theta _{0}}{2},
\\
\Phi &=&-2\left\{ \frac{(1+\lambda \mu )(\lambda +\mu )}{(1+\lambda \mu
)^{2}+(\lambda +\mu )^{2}[q_{1}(\lambda +\mu )+1]^{2}}\right\} \cos ^{-2}%
\frac{\theta _{0}}{2}, \\
\omega &=&\frac{2k}{q_{1}}\left\{ \frac{(1-\mu ^{2})(\lambda
^{2}-1)[1+q_{1}(\lambda +\mu )]}{(1+\lambda \mu )^{2}+(\lambda +\mu
)^{2}[q_{1}^{2}(\lambda +\mu )^{2}-1]}\right\} \cos ^{2}\frac{\theta _{0}}{2}%
.
\end{eqnarray}
More, from (20) with (19), we find for $\gamma $ in (1), 
\begin{equation}
\gamma =\frac{1}{2}\ln \left[ q_{1}^{2}-\frac{(\lambda ^{2}-1)(1-\mu ^{2})}{%
(\lambda +\mu )^{4}}\right] .
\end{equation}
Furthermore, the factor $\cos ^{-2}(\theta _{0}/2)$ in (22) can be absorbed
by a rescaling process of the metric into a conformal metric, like $%
ds_{2}^{2}=\cos ^{2}(\theta _{0}/2)ds^{2}$. Now introducing spherical
coordinates $r$ and $\theta $, through the Boyer-Lindquist transformation 
\cite{Carmeli}, 
\begin{equation}
\lambda =\frac{r-M}{k},\;\;\mu =\cos \theta ,
\end{equation}
into (22) and (24), we obtain asymptotically $r\rightarrow \infty $, 
\begin{eqnarray}
f &\approx &1-2\frac{k}{q_{1}}\frac{1}{r}+O\left( \frac{1}{r^{2}}\right) , \\
\omega &\approx &2\left( \frac{k}{q_{1}}\right) ^{2}\cos ^{2}\frac{\theta
_{0}}{2}\frac{1-\mu ^{2}}{r}+O\left( \frac{1}{r^{2}}\right) .
\end{eqnarray}
We see from (27) and (28) that the solution now has the good asymptotic
behaviour allowing us to interpret the parameters $q_{1}$, $\theta _{0}$ and 
$k$ as 
\begin{equation}
\frac{k}{q_{1}}=M,\;\;\cos ^{2}\frac{\theta _{0}}{2}=\frac{J}{M^{2}}=\frac{a%
}{M},
\end{equation}
where $M$ and $J$ are, respectively, the mass and the angular momentum of
the source, and $a=J/M$ the angular momentum per unit mass. In the Kerr
solution (7) there are two parameters linked by the condition (8). The
asymptotical behaviour of this solution imposes \cite{Kramer} 
\begin{equation}
p=\frac{k}{M},\;\;q=\frac{a}{M},
\end{equation}
and the condition (8) fixes $k$, 
\begin{equation}
k^{2}=M^{2}-a^{2}.
\end{equation}
In our solution, the asymptotical relations (27) and (28) impose (29), but $%
q_{1}$, $\theta _{0}$ and $k$ are arbitrary, as can be seen from (15), (19)
and (26). Of course, it is always possible to compare our parameters to
those of Kerr by putting 
\begin{equation}
\frac{a}{M}=q=\cos ^{2}\frac{\theta _{0}}{2},\;\;\frac{k}{M}=p=q_{1},
\end{equation}
and assuming $0\leq q_{1}\leq 1$. So, we would have also from (8), 
\begin{equation}
q_{1}^{2}+\cos ^{4}\frac{\theta _{0}}{2}=1.
\end{equation}
However, it is not necessary for us to choose (32) and (33). In general, our
solution presents three independent free parameters, $q_{1}$, $\theta _{0}$
and $k$, whereas the Kerr solution presents only one independent parameter,
either $p$ or $q$. Furthermore, imposing (32) and (33), does not reduce our
solution to the Kerr solution. The differences between both solutions are
further studied in the next section. Besides, we note that the solution
(22)-(24) does not belong to the usual Tomimatsu-Sato solutions \cite
{Tomimatsu}.

\section{Horizons, ergospheres and singularities}

The expression (22) can be written like 
\begin{equation}
f=\frac{N}{D}\cos^{-2}\frac{\theta_0}{2},
\end{equation}
with 
\begin{eqnarray}
N=(1+\lambda\mu)^2+(\lambda+\mu)^2[q_1^2(\lambda+\mu)^2-1], \\
D=(1+\lambda\mu)^2+(\lambda+\mu)^2[q_1(\lambda+\mu)+1]^2.
\end{eqnarray}

\subsection{Horizons}

The horizons correspond to the solution of $f=0$ for $\mu =\pm 1$ which is,
from (35), $\lambda =\mp 1$, or from (26), $r_{h}=M\pm k$. These horizons
split into the Cauchy horizon, with radius $r_{ch}=M-k$, and the event
horizon, with radius $r_{eh}=M+k$. These results are satisfactory since, for
any stationary axisymmetric metric, the horizons depend only on the
spacetime symmetries.

\subsection{Ergospheres}

The equation of the ergosphere surfaces, from (34), is $N=0$, and two cases
have to be distinguished.

\subsubsection{$\protect\lambda+\protect\mu=0$}

In this case $N=0$ if, in addition, 
\begin{equation}
1+\lambda\mu=0,
\end{equation}
which imposes two solutions, describing two points, 
\begin{eqnarray}
\lambda=-1, \;\; \mu=1, \\
\lambda=1, \;\; \mu=-1.
\end{eqnarray}
These points, (38) and (39), are the intersections of the $z$ axis with the
event horizon, $r=r_{eh}=M+k$, and Cauchy horizon, $r=r_{ch}=M-k$,
respectively, belonging to the ergospheres. It has to be noted that (38) and
(39) produce, from (36), $D=0$ as well, hence there is an indetermination
for the ratio $N/D$. This indetermination can be raised by studying the
limits $\lambda\rightarrow\pm 1$ which produce 
\begin{equation}
\lim_{\lambda\rightarrow -1}f(\mu=1)=\lim_{\lambda\rightarrow 1}f(\mu=-1)=0.
\end{equation}
The limits (40) are finite and zero, hence these points belong to the
ergospheres.

\subsubsection{$\protect\lambda+\protect\mu\neq 0$}

In this case, $(\lambda+\mu)^2$ can be factorized in (35), and the equation
for $N=0$ becomes 
\begin{equation}
\left(\frac{1+\lambda\mu}{\lambda+\mu}\right)^2+q_1^2(\lambda+\mu)^2=1,
\end{equation}
which is the equation for the ergosphere surfaces. It is a fourth degree
surface, and for the representation of this surface, it is useful to express
it through a parametric representation with the help of a parameter $\tau$,
such that, 
\begin{equation}
\left(\frac{1+\lambda\mu}{\lambda+\mu}\right)^2=\cos^2\tau, \;\;
q_1^2(\lambda+\mu)^2=\sin^2\tau.
\end{equation}
We can see from (42) that it is a bounded closed surface for any value of
the $q_1$ parameter. We have plotted some curves, which are intersections of
this surface by the meridian plane $\phi=0$, for different values of the
parameter $1\geq q_1>0$, as shown in figs. 1-10. These curves present the
following interesting features.

\begin{itemize}

\item  When $q_{1}\rightarrow 0$ the aspect of the ergospheres and horizons tends
towards the aspect of the Kerr extreme black hole (e.g. see fig.4 of \cite
{Gariel}), as shown in fig.1.

\item  When $q_{1}$ increases its value the aspect of the ergospheres remarkably
differs from this of a Kerr black hole, as shown in figs. 2-5. Specially, we
can notice, the surface of the exterior ergosphere becomes double,
presenting some thickness being a two-sheet torus. It is the same for the
interior ergosphere.

\item  For a defined value of $q_1$, near $q_1\approx 0.5$, the exterior
ergosphere opens itself on the axis $\mu=0$ ($\theta=\pi/2$), as shown in
figs. 6-8. Then the event-horizon becomes naked in a certain angular
aperture, whereas the Kerr event-horizon is always dressed by the exterior
surface of the ergosphere. Thus, on this spatial portion, the Penrose
process \cite{Penrose} is no longer able to take place. This special
topology of the ergosphere indicates, also here, a difference with the Kerr
metric.

\item  The evolution of the interior part of the ergosphere, for increasing
values of $q_{1}$, looks intricate, with, particularly, the advent from the
center of a new curve, as shown in figs. 3-4, with a four-leaved clover
shape, which grows up until to pass beyond the Cauchy horizon, as shown in
figs. 8-9, which of course vanishes when $q_{1}=1$ ($M=k$), as shown in fig.
10. This complicated behaviour presents, also here, an important difference
with the Kerr metric, because in this last case, the Cauchy horizon always
covers the interior ergosphere.
\end{itemize}

Figs. 1--10 show the parametric plots of the curves describing the
intersections of the interior and exterior ergospheres,defined by eq.(4.1),
with the meridian plane $\phi =0$ for different values of the parameter $%
q_{1}$ in the range $\left[ 10^{-2},1\right] $. The vertical axis is z.\ The
ergospheres are the axisymmetric surfaces which can be generated by rotation
of the curves around the z-axis. The event-horizon and Cauchy-horizon are
also represented (circles of radius $r_{eh}=M+k$, $\ r_{ch}=M-k$,
respectively). The mass M has been fixed to the value $M=4$.\ $k$ is given
by eq.(29).

\subsection{Singularities}

The singularities correspond, when they exist, to curves or surfaces defined
by $D=0$ from (36). We see that $D$ is a sum of squares and it can vanish
only in two cases:

\subsubsection{$1+\protect\lambda\protect\mu=0$ and $\protect\lambda+\protect%
\mu=0$}

This system of equations is the same as studied in 4.2.1, and corresponds to
the two points (38) and (39) of the horizons where $N=0$. Since, after
raising the indetermination of the ratio $N/D$, the limit (40) is finite and
zero, these two points are not singular.

\subsubsection{$1+\protect\lambda\protect\mu=0$ and $q_1(\protect\lambda+%
\protect\mu)=-1$}

Or, equivalently, 
\begin{eqnarray}
\lambda=-\frac{1}{\mu}, \\
q_1\mu^2+\mu-q_1=0.
\end{eqnarray}
The polynomial (44) always has two roots, 
\begin{equation}
\mu_{\pm}=\frac{-1\pm\sqrt{\Delta}}{2q_1}, \;\; \Delta=1+4q_1^2,
\end{equation}
that gives the two solutions, 
\begin{equation}
\mu_+=\frac{-1+\sqrt{\Delta}}{2q_1}, \;\; \lambda_+=-\frac{1}{\mu_+},
\end{equation}
and 
\begin{equation}
\mu_-=-\frac{1+\sqrt{\Delta}}{2q_1}, \;\; \lambda_-=-\frac{1}{\mu_-}.
\end{equation}
The first solution, (46), produces $0\leq\mu_+\leq 1$, while the second,
(47), produces $|\mu_-|>1$, hence it has to be rejected. From (46) with
(26), we have 
\begin{equation}
r_+=M\left(1+\frac{2q_1^2}{1-\sqrt{\Delta}}\right),
\end{equation}
which gives $r_+<r_{ch}=M-k$, hence the two ring singularities (48), which
are the solutions for $\mu_+=\cos(\pm\theta_+)$, are inside the Cauchy
horizon and so, {\it a fortiori}, inside the event horizon. There are no
naked singularities.

\section{Conclusion}

It has been proposed a new axially symmetrical stationary vacuum solution
(15) of Ernst equation. Unfortunately, this solution does not satisfy the
aymptotical flatness. Only after performing an Ehlers transformation and an
unitary transformation, the solution (22)-(24) achieves the appropriate
physical asymptotical flatness. Three arbitrary parameters were introduced
in this process and interpreted from the asymptotical properties of the
solution related to the total mass of its source and its angular momentum.
We did not succeed to obtain the Kerr limit of the solution, however we know
that it does exist because of the uniqueness theorem, since this new
solution has asymptotical flatness and does not present naked singularities.
One of the parameters introduced shapes the ergosphere demonstrating big
differences to the Kerr solution. When this parameter vanishes the solution
becomes the extreme Kerr black hole. We might conjecture that the solution
(22)-(24) represents a distorted stationary black hole as obtained in the
static case \cite{Geroch}.

\section*{Acknowledgments}

The authors wish to thank R.Colistete Jr. for his valuable help discussing
and formatting this article.

\newpage

\begin{figure}
\begin{center}
\includegraphics[trim=0in 0.10in 0in 0.10in, scale=0.63]{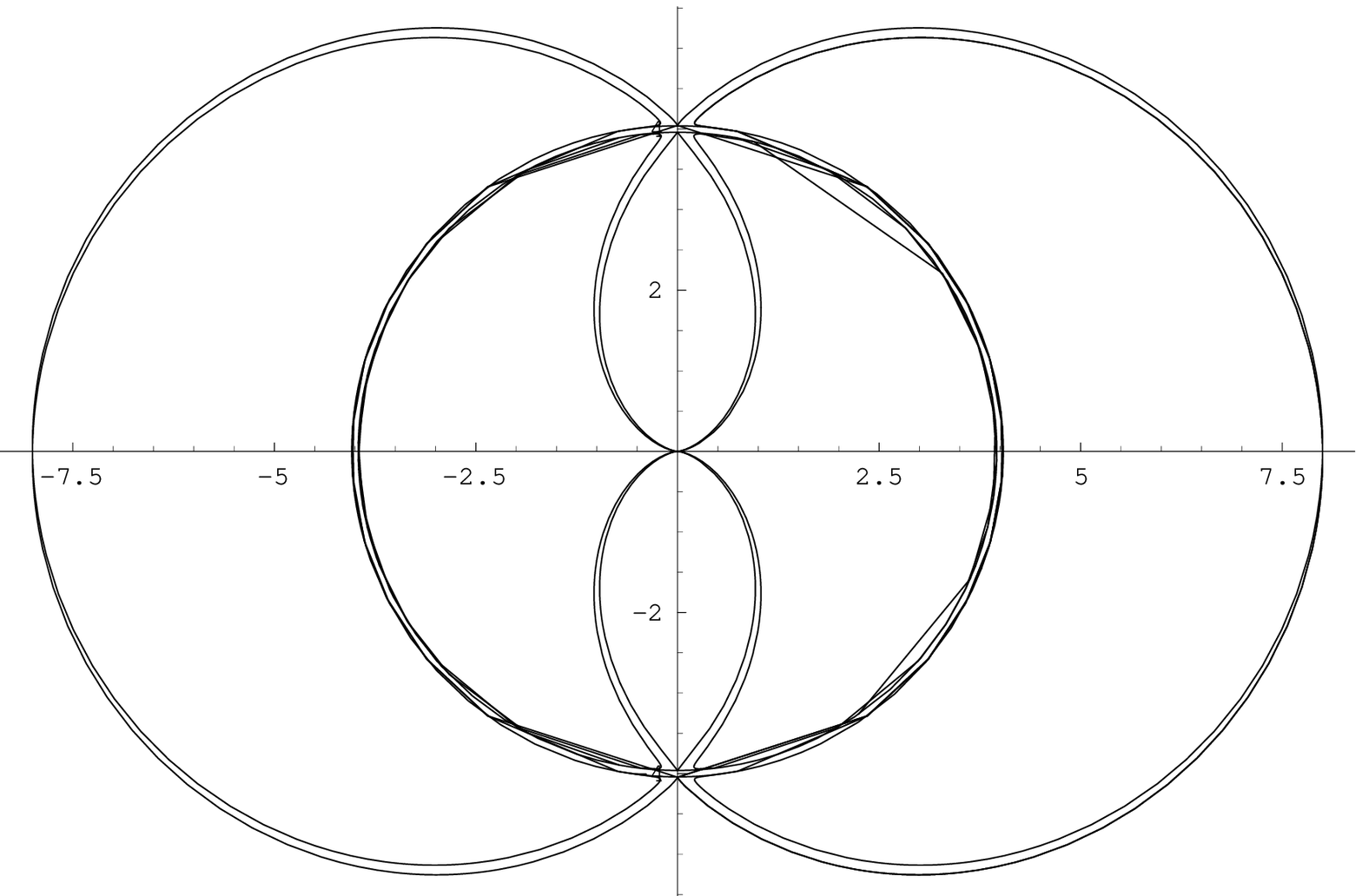}
\end{center}
\caption{$q_{1}=0.01$, $r_{eh}=4.04$, $r_{ch}=3.96$.}
\end{figure}

\begin{figure}
\begin{center}
\includegraphics[trim=0in 0.10in 0in 0.10in, scale=0.63]{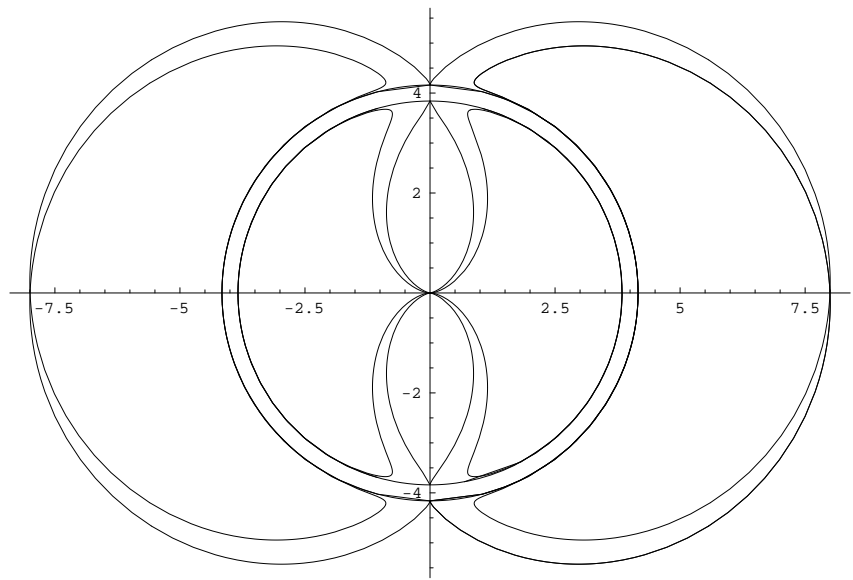}
\end{center}
\caption{$q_{1}=0.04$, $r_{eh}=4.16$, $r_{ch}=3.84$.}
\end{figure}

\begin{figure}
\begin{center}
\includegraphics[trim=0in 0.15in 0in 0.15in, scale=0.57]{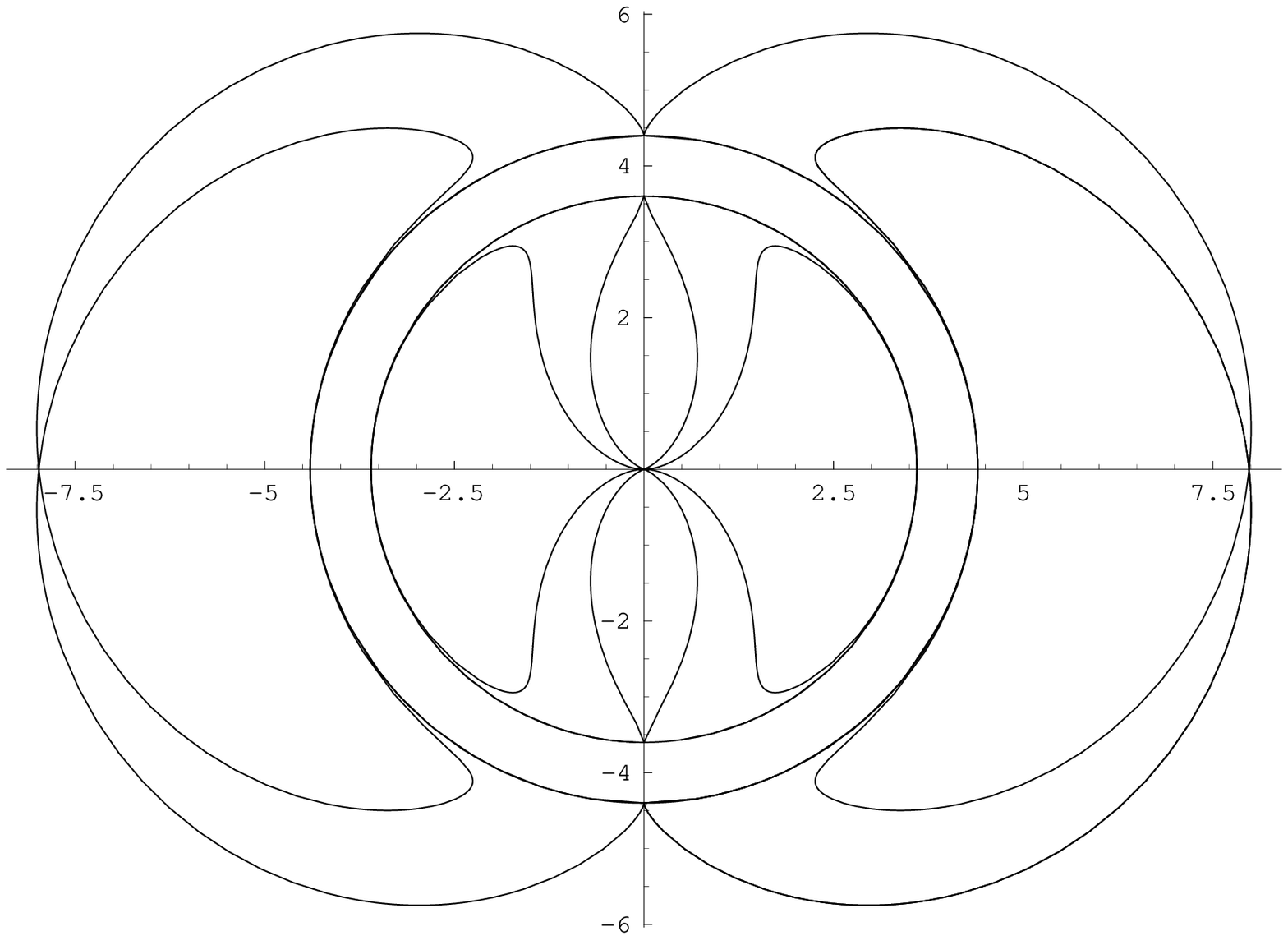}
\end{center}
\caption{$q_1 = 0.1$, $r_{eh} = 4.4$, $r_{ch} = 3.6$.}
\end{figure}

\begin{figure}
\begin{center}
\includegraphics[trim=0in 0.10in 0in 0.10in, scale=0.57]{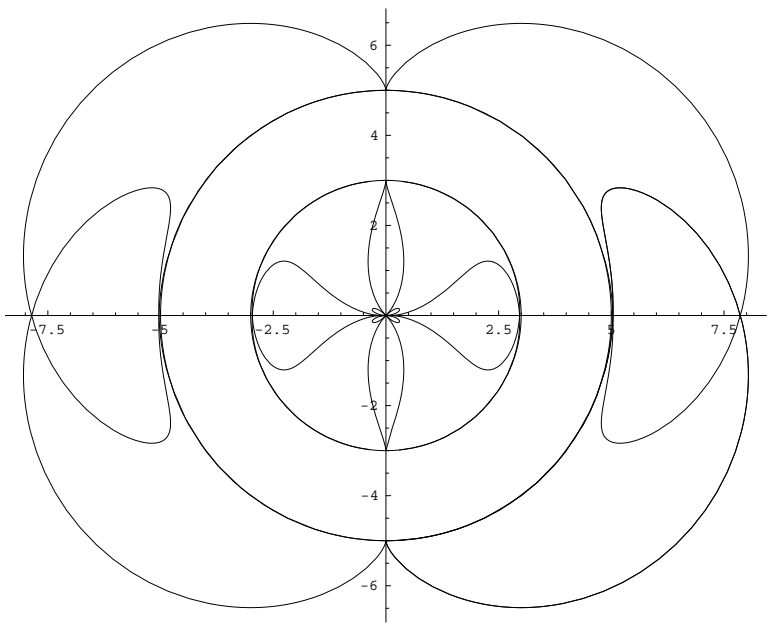}
\end{center}
\caption{$q_1 = 0.25$, $r_{eh} = 5$, $r_{ch} = 3$.}
\end{figure}

\begin{figure}
\begin{center}
\includegraphics[trim=0in 0.10in 0in 0.10in, scale=0.47]{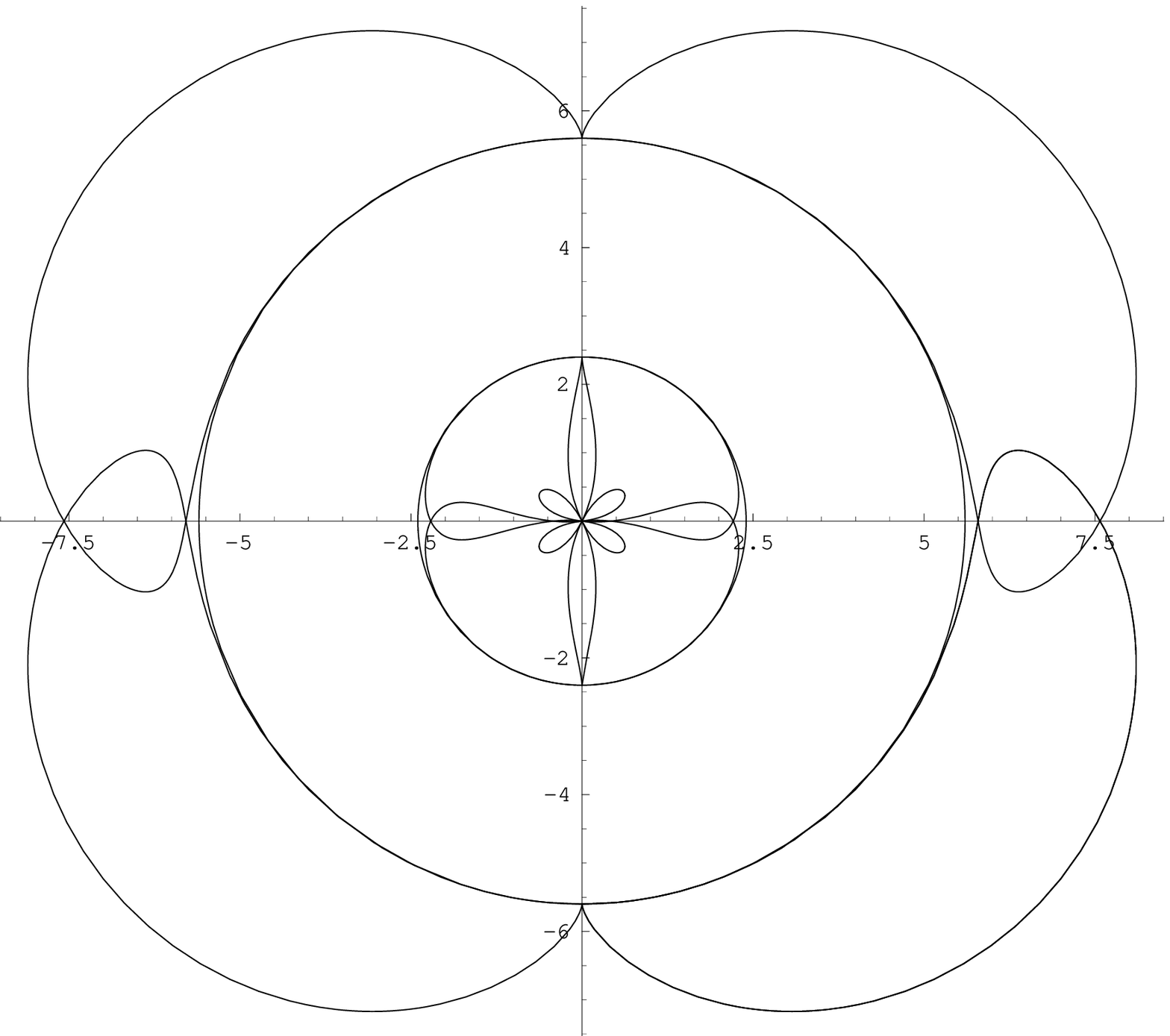}
\end{center}
\caption{$q_1 = 0.4$, $r_{eh} = 5.6$, $r_{ch} = 2.4$.}
\end{figure}

\begin{figure}
\begin{center}
\includegraphics[trim=0in 0.15in 0in 0.15in, scale=0.47]{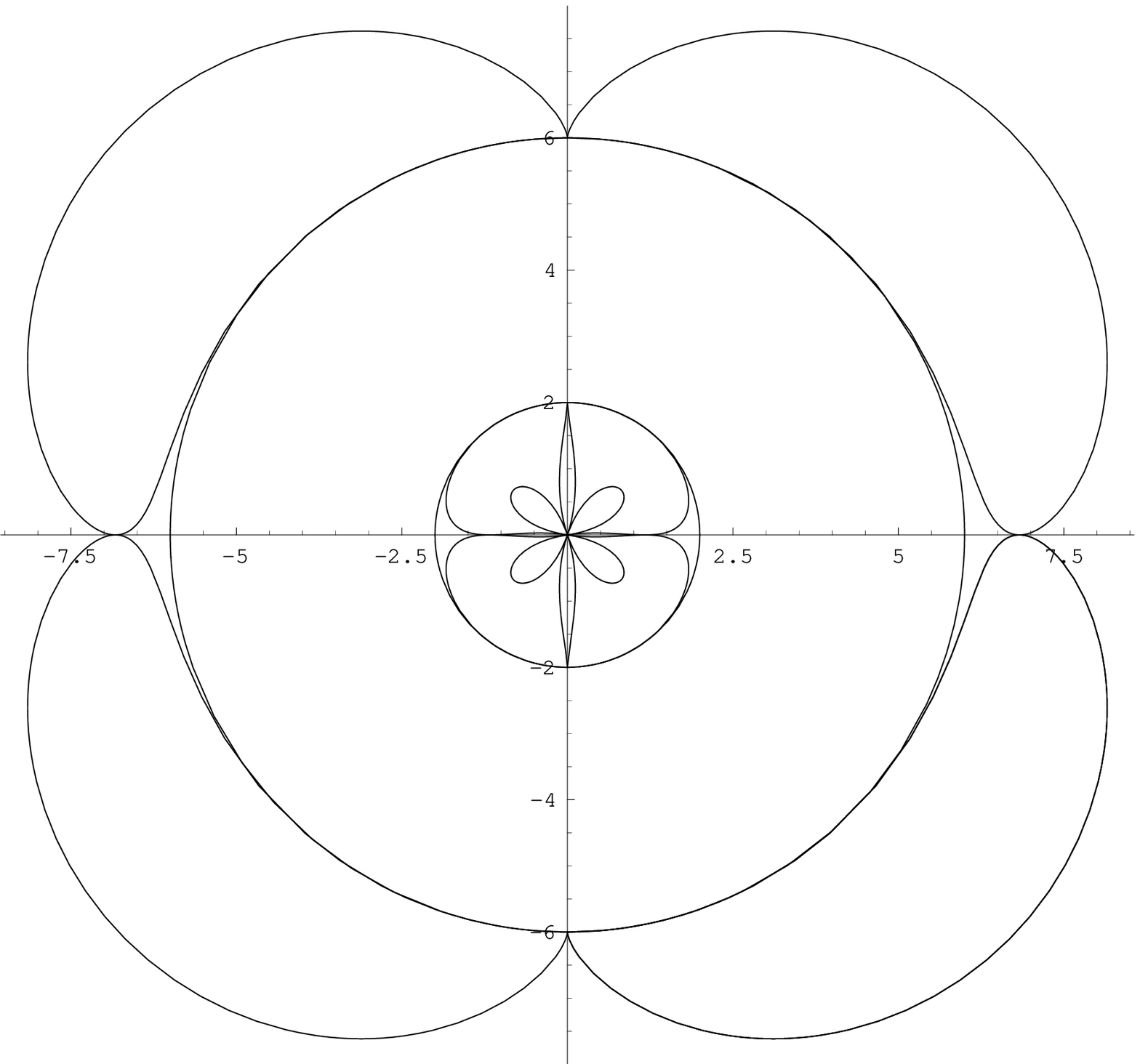}
\end{center}
\caption{$q_1 = 0.5$, $r_{eh} = 6$, $r_{ch} = 2$.}
\end{figure}

\begin{figure}
\begin{center}
\includegraphics[trim=0in 0.10in 0in 0.10in, scale=0.42]{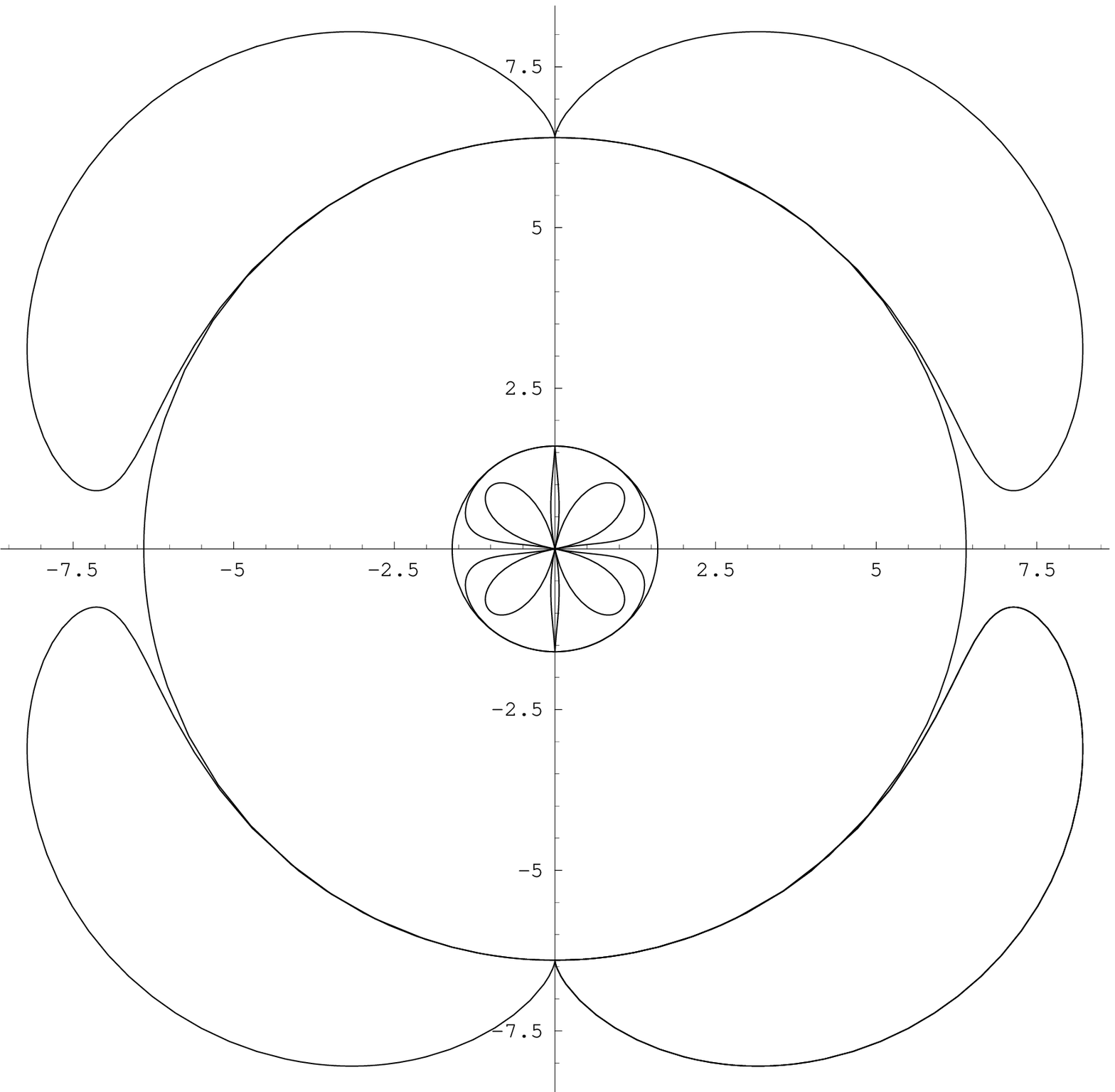}
\end{center}
\caption{$q_1 = 0.6$, $r_{eh} = 6.4$, $r_{ch} = 1.6$.}
\end{figure}

\begin{figure}
\begin{center}
\includegraphics[trim=0in 0.15in 0in 0.15in, scale=0.42]{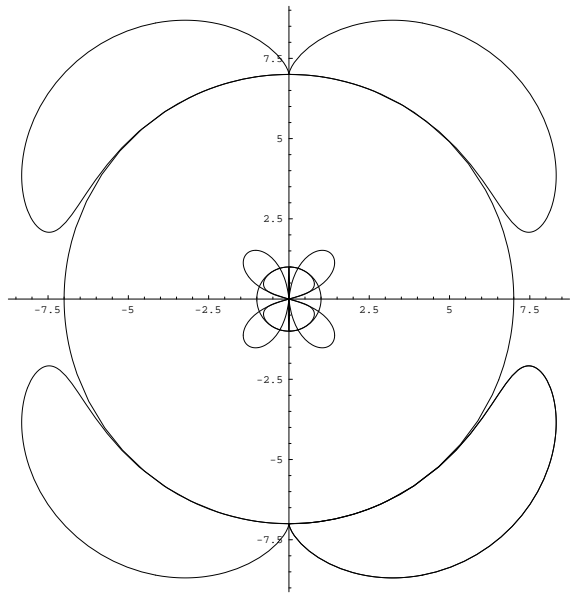}
\end{center}
\caption{$q_1 = 0.75$, $r_{eh} = 7$, $r_{ch} = 1$.}
\end{figure}

\begin{figure}
\begin{center}
\includegraphics[trim=0in 0.15in 0in 0.15in, scale=0.37]{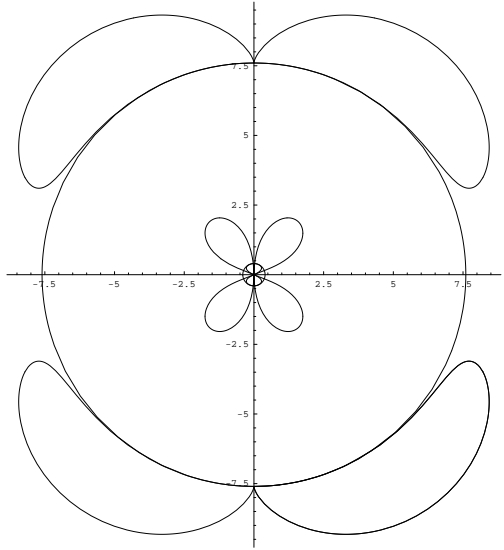}
\end{center}
\caption{$q_1 = 0.9$, $r_{eh} = 7.6$, $r_{ch} = 0.4$.}
\end{figure}

\begin{figure}
\begin{center}
\includegraphics[scale=0.37]{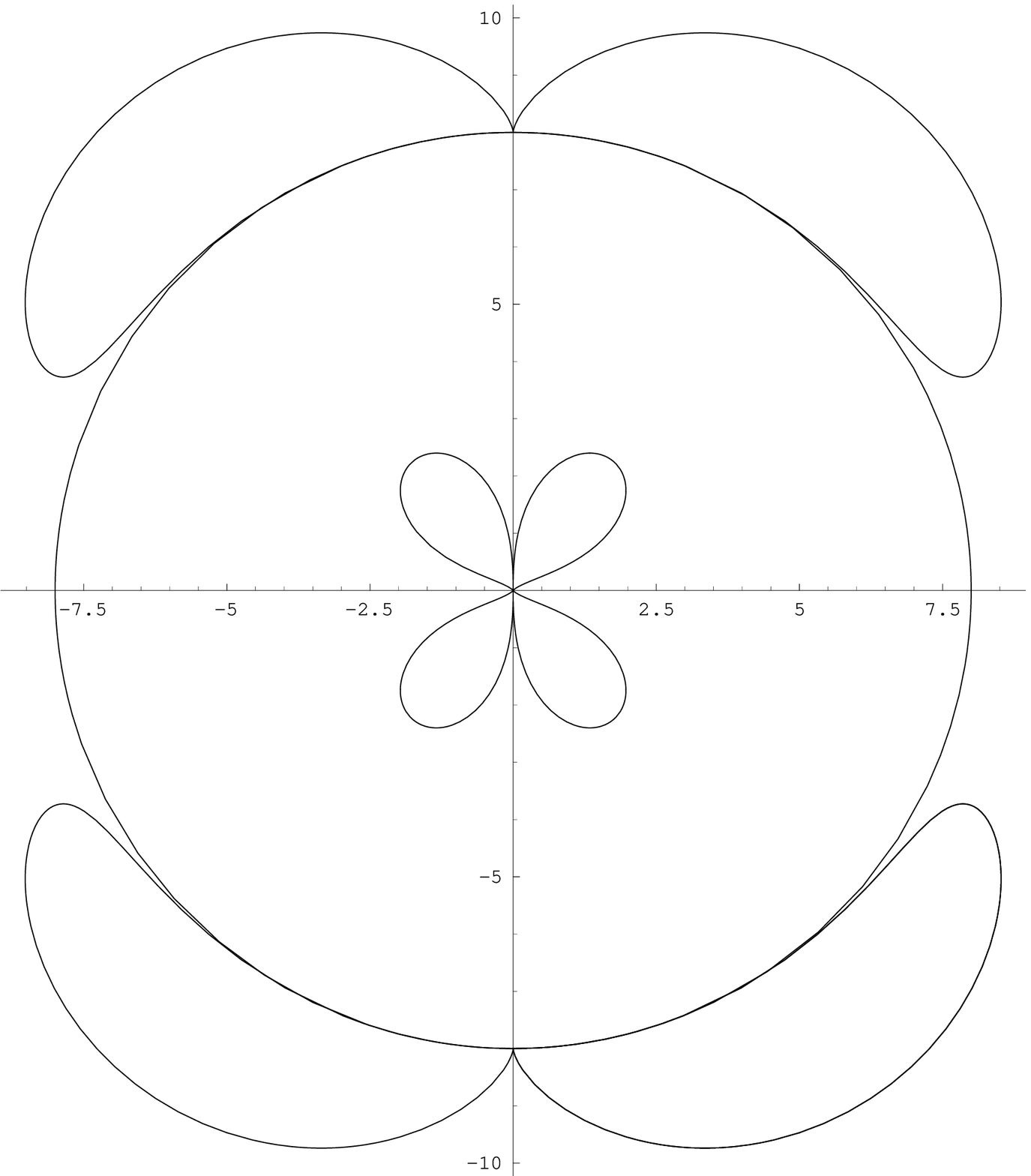}
\end{center}
\caption{$q_1 = 1$, $r_{eh} = 8$, $r_{ch} = 0$.}
\end{figure}


\begin{thebibliography}{9}
\bibitem{Carter}  Carter B 1971 {\it Phys. Rev. Lett.} {\bf 26}, 331;
Robinson D C 1975 {\it Phys. Rev. Letters} {\bf 34}, 905.

\bibitem{Penrose}  Penrose R and Floyd R M 1971 {\it Nature Phys. Sci.} {\bf %
229}, 177.

\bibitem{Ernst}  Ernst F J 1968 {\it Phys. Rev.} {\bf 167}, 1175.

\bibitem{Tomimatsu}  Tomimatsu A and Sato H 1972 {\it Phys. Rev. Lett.} {\bf %
29}, 1344; 1973 {\it Prog. Theor. Phys.} {\bf 50}, 95.

\bibitem{Chandrasekhar}  Chandrasekhar S 1983 {\it The Mathematical Theory
of Black Holes} (Oxford University Press, Oxford) p. 285.

\bibitem{Carmeli}  Carmeli M 1982 {\it Classical Fields: General Relativity
and Gauge Theory} (John Wiley and Sons, New York) p. 387.

\bibitem{Kramer}  Kramer D, Stephani H, Herlt E and MacCallum M 1980 {\it %
Exact Solutions of Einstein's Field Equations} (Cambridge University Press,
Cambridge) p. 206.

\bibitem{Gariel}  Gariel J, Marcilhacy G, \ Santos NO and Colistete R (2000).%
{\it Parametrization of singularities of the Demiamski-Newman spacetimes},
gr-qc/0012004.

\bibitem{Geroch}  Geroch R and Hartle J B 1982 {\it J. Math. Phys.} {\bf 23}%
, 680.
\end{thebibliography}
\end{document}